\documentclass[prb,onecolumn,showpacs,superscriptaddress,preprintnumbers,amssymb]{revtex4}
\usepackage{graphicx}
\usepackage{dcolumn}
\usepackage{bm}
\usepackage{wasysym}
\usepackage{color}

\newcommand{\beq}{\begin{equation}}
\newcommand{\eeq}{\end{equation}}
\newcommand{\beqn}{\begin{eqnarray}}
\newcommand{\eeqn}{\end{eqnarray}}
\newcommand{\nn}{\nonumber\\}

\begin{document}
\title{Instability of the non-Fermi liquid state of the Sachdev-Ye-Kitaev Model }

\author{Zhen Bi}
\affiliation{Department of Physics, University of California,
Santa Barbara, CA 93106, USA}
\author{Chao-Ming Jian}
\affiliation{Kavli Institute of Theoretical Physics, Santa
Barbara, CA 93106, USA}
\author{Yi-Zhuang You}
\affiliation{Department of Physics, Harvard University, Cambridge,
MA 02138, USA}
\author{Kelly Ann Pawlak}
\affiliation{Department of Physics, University of California,
Santa Barbara, CA 93106, USA}
\author{Cenke Xu}
\affiliation{Department of Physics, University of California,
Santa Barbara, CA 93106, USA}

\date{\today}
\begin{abstract}

We study a series of perturbations on the Sachdev-Ye-Kitaev (SYK)
model. We show that the maximal chaotic non-Fermi liquid phase
described by the ordinary $q = 4$ SYK model has marginally
relevant/irrelevant (depending on the sign of the coupling
constants) four-fermion perturbations allowed by symmetry.
Changing the sign of one of these four-fermion perturbations leads
to a continuous chaotic-nonchaotic quantum phase transition of the
system accompanied by a spontaneous time-reversal symmetry
breaking. Starting with the SYK$_q$ model with a $q-$fermion
interaction, similar perturbations can lead to a series of new
fixed points with continuously varying exponents.

\end{abstract}
\maketitle

\section{Introduction}

Non-Fermi liquids usually occur at quantum critical points of
itinerant electron systems~\cite{hertz,millis,nfl}. Strong
correlation and quantum critical fluctuation often make it
challenging to study the non-fermi liquids through the standard
diagrammatic approach, and various expansion methods have been
developed for that
purpose~\cite{lee2009,senthilnfl,maxnfl1,maxnfl2,leenfl}.
Fortunately, there exist some exactly soluble models for non-Fermi
liquid states which do not rely on perturbation theory. In 1993,
Sachdev and Ye constructed one such example in
$(0+1)d$~\cite{SachdevYe1993}, which was reintroduced in a
modified version lately by Kitaev~\cite{Kitaev2015}. This model is
now known as the Sachdev-Ye-Kitaev (SYK) model. The SYK model is a
$(0+1)d$ system that consists of $N$ Majorana fermions with
$q$-fermion random interactions. When $q=2$, the model is simply
$N$ Majorana fermions with only random hopping terms, which can be
solved completely using the random matrix theory. The $q = 4$ SYK
model (hereafter labelled as SYK$_4$ model) is most thoroughly
studied. Its Hamiltonian is given by \beq H_{\text{SYK}_4} =
\sum_{ijkl} \frac{J_{ijkl}}{4!}\chi_i\chi_j\chi_k\chi_l, \eeq
where $\chi_{i,j,k,l}$ are Majorana fermion operators with index
$i,j,k,l=1 \cdots N$, and $J_{ijkl}$ is a fully anti-symmetric
tensor whose each entry is drawn from a Gaussian distribution with
zero mean and variance $\overline{J_{ijkl}^2} = 3! J_4^2/N^3$.
With large $N$ and low temperature, the SYK$_4$ model can be
solved exactly via saddle point equations and exhibits an emergent
conformal symmetry. The scaling dimension of the Fermion operator
is $\Delta_f = 1/4$, which suggests a non-Fermi liquid behavior
without quasi-particle excitations
\cite{Kitaev2015,MaldacenaStanford2016}.

Furthermore, the exact solution also suggests that the SYK$_4$
model is maximally chaotic, in the sense that its Lyapunov
exponent~\cite{Kitaev2015,MaldacenaStanford2016}, a measure of
quantum chaos, saturates the universal upper bound established in
Ref.~\onlinecite{MSSbound}. The saturation of the universal upper
bound is also a feature of black holes. In fact, the exact
solution also indicates that the SYK$_4$ model should indeed be
holographically dual to a gravity
theory~\cite{Sachdev2010,Sachdev2015,Polchinski2016,jensen2016,verlinde2016,MaldacenaStanford2016,MSY2016}.
All SYK$_q$ models share the properties such as maximally chaotic
non-Fermi liquid ground states (for $q>2$), emergent conformal
symmetry at large-$N$~\footnote{As was pointed out in
Ref.~\onlinecite{MaldacenaStanford2016}, rigorously speaking the
full reparametrization symmetry of this model is broken both
spontaneously and explicitly, thus the conformal symmetry is
approximate.}, etc.
Many other aspects of the SYK model, including the numerical
simulations, generalizations to models with higher symmetry, and
higher dimensions, have been investigated recently
\cite{Georges2001,you2016,
Fu2016,Cotler2016,antonio2016,Sachdev2015,FuSUSY2016,Gu2016,
Davison2016, Witten2016, Klebanov2016, Verlinde2017,
Banerjee2016,gross2016,krishnan2016}.

One peculiar feature of the SYK$_q$ model with $q>2$ is that, in
the large $N$ limit, the chaotic non-Fermi liquids all have finite
entropy density even when the temperature approaches
zero~\cite{Georges2001,Sachdev2015,Kitaev2015,MaldacenaStanford2016}.
One might conjecture directly that the system has instabilities
towards states with lower (or zero) zero-temperature entropy
density upon perturbations. Indeed, in experimental systems, the
non-Fermi liquid state at a quantum critical point is usually
buried in a dome of ordered phase with spontaneous symmetry
breaking at low temperature~\cite{maxnfl}. One usual scenario is
the emergence of a superconducting dome around the quantum
critical point, which occurs in cuprates, pnictides
superconductors, and also some heavy fermion systems. Thus it is
meaningful to ask whether the SYK$_q$ model, especially the
SYK$_4$ model is instable against spontaneous symmetry breaking.
Or in other words, the SYK$_4$ model could be the parent state of
ordered phases at the infrared~\footnote{Here we use the standard
Landau-Ginzburg's definition of an ordered phase: an order means
some symmetry of the system is spontaneously broken, or in other
words, an order parameter that transforms nontrivially under the
symmetry acquires a long range correlation.}.

In this paper, we study a class of perturbations on the SYK$_q$
models. We will concentrate mostly on the case with $q = 4$.
Obviously, the non-Fermi liquid at the SYK$_4$ fixed point will be
unstable against the SYK$_2$ perturbation. However, the SYK$_4$
has a time-reversal symmetry, under which all fermion bilinears
are odd. The time-reversal symmetry $\mathcal{T}$ forbids
perturbations like the SYK$_2$ term. Thus we only consider
four-fermion terms which are symmetric under $\mathcal{T}$. As we
will show, the non-Fermi liquid SYK$_4$ model is instable against
a series of four-fermion interactions that preserve all the
symmetries, and the system flows to a state with spontaneous
breaking of $\mathcal{T}$.

A similar analysis can be generalized to the SYK$_q$ non-Fermi
liquid with $q > 4$ perturbed by the four-Fermion interactions we
design. Interestingly, the four fermion interactions can drive the
SYK$_q$ model to a series of new stable fixed points with
conformal symmetry.


\section{A perturbed $q=4$ SYK model}

\begin{figure}[tbp]
\begin{center}
\includegraphics[width=240pt]{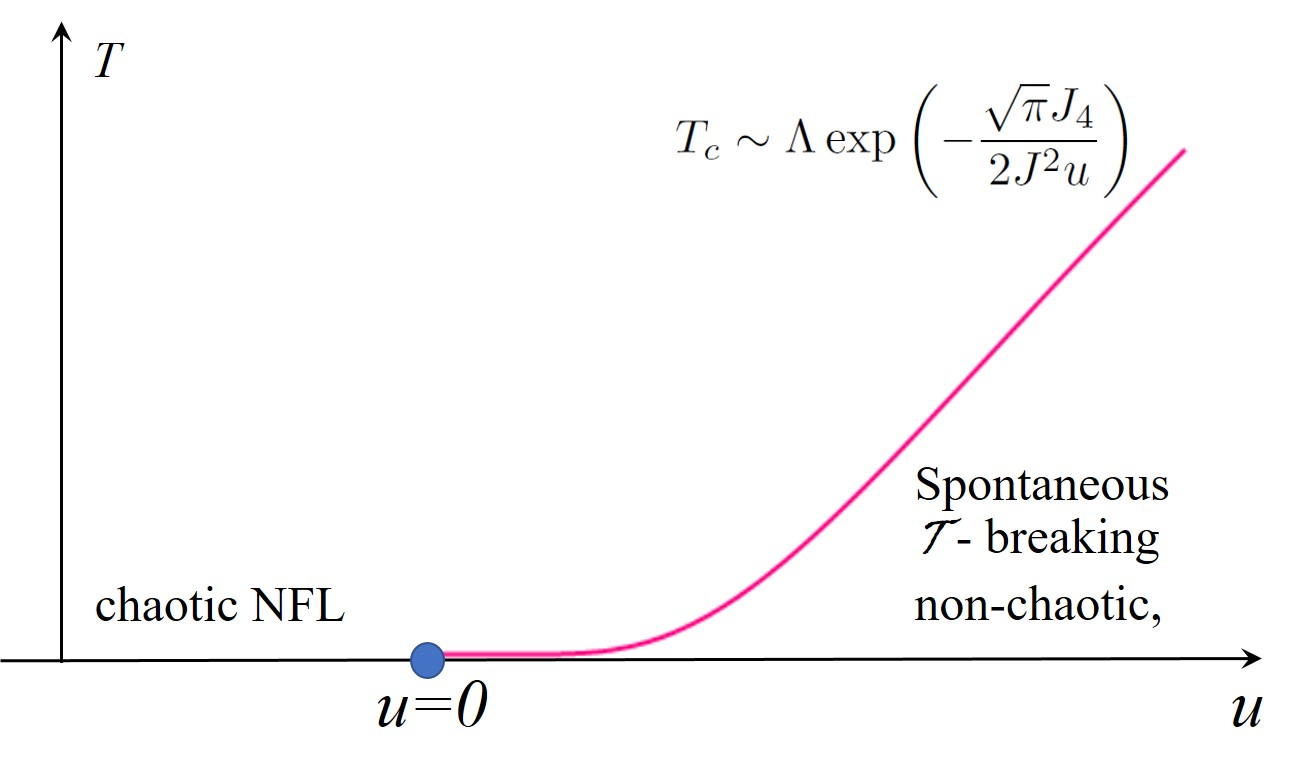}
\caption{The phase diagram of Eq.~\ref{cluster}.} \label{pd}
\end{center}
\end{figure}

The goal of the first section is to study the following
generalized SYK model: \beq \label{cluster} H
=\frac{J_{ijkl}}{4!}\chi_i\chi_j\chi_k\chi_l + \frac{u}{2}
C_{ij}C_{kl}\chi_i\chi_j\chi_k\chi_l \eeq Both $J_{ijkl}$ and
$C_{ij}$ are anti-symmetric random tensors drawn from a gaussian
distribution. We choose the following normalization for $J_{ijkl}$
and $C_{ij}$: \beqn \label{norm} \overline{J_{ijkl}}=0,&\ \ \
N^3\overline{J_{ijkl}^2} = 3!J_4^2 \nn \overline{C_{ij}}=0,&\ \ \
N^2\overline{C_{ij}C_{kl}} = J^2 (\delta_{ik}\delta_{jl} -
\delta_{il}\delta_{jk}). \eeqn Note that $J_4$ has the dimension
of energy, while $J$ has the dimension of (energy)$^{1/2}$. The
results of this section is summarized in phase diagram
Fig.~\ref{pd}.

The two terms in Eq.~\ref{cluster} have the same symmetry: the
time-reversal symmetry $\mathcal{T}$ which acts as $\chi_j
\rightarrow \chi_j$, $i \rightarrow -i$ (it is the same
time-reversal symmetry of the boundary states of the topological
superconductor in the BDI
class~\cite{kitaevclass,ludwigclass1,ludwigclass2}), and a
statistical O($N$) symmetry. We will demonstrate that, by tuning
$u$ from negative to positive, the system goes through a
continuous phase transition from a chaotic phase to a nonchaotic
phase. The critical properties of this transition are analogous to
that of the Kosterlitz-Thouless transition, with exponent $\nu = +
\infty$.

\subsection{The $u-$term}

Before we study Eq.~\ref{cluster}, let us start with the
Hamiltonian with only the second term: \beqn H^\prime =
\frac{u}{2} C_{ij}C_{kl}\chi_i\chi_j\chi_k\chi_l. \eeqn This
Hamiltonian can be written as $H^\prime = - u \hat{b}^2/2 $, with
$\hat{b} = i C_{jk}\chi_j \chi_k$. Since $\hat{b}$ commutes with
$H^\prime$, it is a conserved quantity. Thus every eigenstate of
$H^\prime$ is an eigenstate of $\hat{b}$ with eigenvalue $b$. When
$u > 0$, the ground state of $H^\prime$ has the maximum eigenvalue
of $\hat{b}$.

Now we can view $\hat{b}$ as a quadratic fermion Hamiltonian with
random hopping. To maximize $\hat{b}$, the system fills all the
negative (or positive) eigenvalues of the single fermion energy
level $\varepsilon_l$, and $\mathrm{Max}[ |b| ] = |\sum
\varepsilon_l|$ with $\varepsilon_l < 0$.

The single particle energy levels $\varepsilon_l$ are the
eigenvalues of the random Hermitian matrix $iC$. Based on the
semi-circle law, the average number of eigenvalues of $iC$ in
$(\varepsilon, \varepsilon + d\varepsilon)$ is given by
$\rho(\varepsilon) d\varepsilon$ with \beq \rho(\varepsilon) =
\frac{N^2}{2\pi J^2} \sqrt{\frac{4J^2}{N} - \varepsilon^2}. \eeq
Then we can obtain the average value of $\mathrm{Max}[ |b| ] $ as
\beq \mathrm{Max}[ |b| ]  = \Bigg|\int_{\varepsilon<0 }
d\varepsilon \varepsilon \rho(\varepsilon)  \Bigg| = \frac{4 J
N^{\frac{1}{2}}}{3\pi}. \eeq Therefore, the average ground state
energy of $H^\prime$ is $E_0(H^\prime) = - \frac{16uJ^2
N}{9\pi^2}$. Thus just like the ordinary SYK model, $H^\prime$
normalized as in Eq.~\ref{norm} is an order-$N$ term.

For $u < 0$, all states with $b = 0$ are ground states, and $b =
0$ is a very ``loose" condition. We will argue that $H^\prime$
with $u < 0$ behaves like a completely free system with zero
Hamiltonian. The (many-body) spectrum of $\hat{b}$ is given by $b
= \sum_{\varepsilon_l > 0} \varepsilon_l n_l$, where the
occupation number $n_l = \pm 1$. This expression of $b$ is similar
to an $\frac{N}{2}$-step random walk centered around $0$. The
distribution of $b$ should therefore be Gaussian. The standard
deviation $\sigma_b$ of this ``random walk" is given by \beq
\sigma_b^2 = \sum_{\varepsilon_l > 0} \varepsilon_l^2 =
\frac{1}{2} \text{Tr} \left((iC)^\dag (iC)\right) = \sum_{i<j}
|C_{ij}|^2 =  \frac{N-1}{2N}J^2. \eeq The (many-body) density of
states of $\hat{b}$ can be then approximated by \beq \rho(b) =
2^{\frac{N}{2}}\sqrt{\frac{ N}{\pi (N-1) J^2}} e^{-\frac{N
b^2}{(N-1)J^2}}, \eeq namely the number of eigenvalues of
$\hat{b}$ in $(b,b +db)$ is given by $\rho(b) db$. The expression
$\rho(b)$ of the density of states $\hat{b}$ is most accurate near
$b=0$, which is exactly the region of interest when $u<0$. We can
now calculate the partition function \beq \mathcal{Z} = \int d b
\rho(b) e^{\beta u b^2} = 2^{\frac{N}{2}} \frac{1}{\sqrt{1+\beta
|u| \frac{N-1}{N} J^2}}. \eeq The entropy density $\mathcal{S}$
can be written as $\mathcal{S} = \frac{1}{N} \left( \log
\mathcal{Z} - \beta \frac{\partial}{\partial \beta} \log
\mathcal{Z} \right)$. Interestingly, we notice that, for any fixed
$\beta$, \beq \lim_{N\rightarrow \infty} \mathcal{S} =
\frac{1}{2}\log 2. \eeq Therefore, if we take the large $N$ limit
first before we take $\beta \rightarrow \infty$, we will conclude
that the ``ground state" entropy density is given by $\frac{1}{2}
\log 2$. Such an entropy density is exactly the same as the system
with zero Hamiltonian. Therefore, we argue that the system with
$u<0$ behaves like a completely free system with zero Hamiltonian.
Using the partition function, we can also calculate the specific
heat of $H^\prime$ with $u<0$: \beq c_v = -\beta \frac{d S}{d
\beta} = \frac{1}{2N} \left( \frac{\frac{N-1}{N}|u|J^2 \beta
}{1+\frac{N-1}{N}|u|J^2 \beta}\right)^2. \eeq

\subsection{Renormalization Group of $u$}

\begin{figure}[tbp]
\begin{center}
\includegraphics[width=260pt]{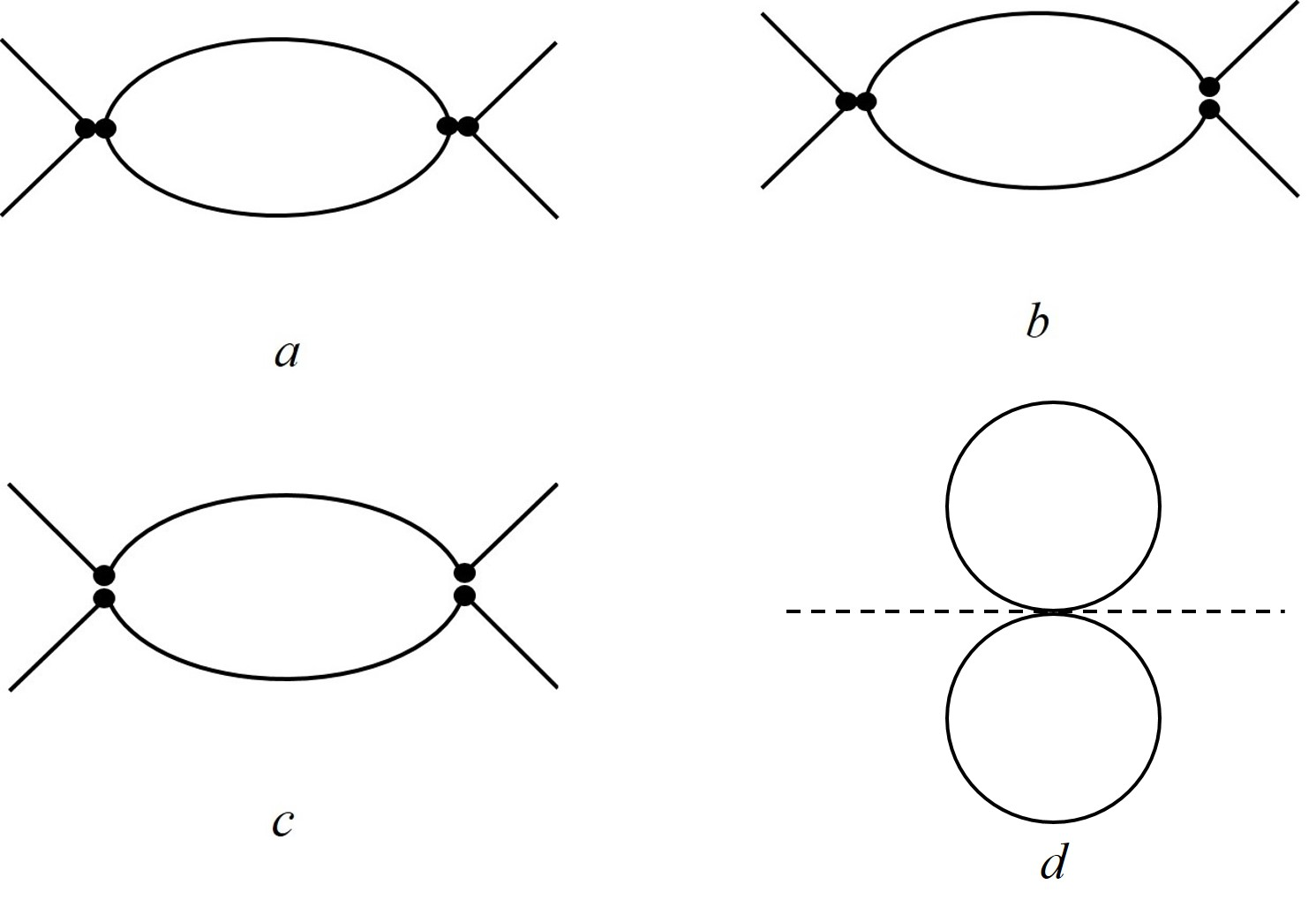}
\caption{($a$), ($b$), ($c$), the diagrams that we consider for
the leading order RG for the coupling constant $u$ in
Eq.~\ref{cluster}. Only diagram ($a$) contributes in the large$-N$
limit. ($d$), the leading order RG for $u$ in Eq.~\ref{replica},
which is equivalent to ($a$), the solid and dashed lines are
fermion and boson Green's functions.} \label{RG}
\end{center}
\end{figure}

When $u$ is treated as a perturbation in Eq.~\ref{cluster}, power
counting indicates that it is a marginal perturbation at the
SYK$_4$ fixed point. Now we perform a perturbative renormalization
group calculation for $u$. We evaluate the fermion Green's
function at the SYK$_4$ fixed point: \beqn G(\tau) =
\left(\frac{1}{4\pi}\right)^{1/4}\frac{\mathrm{sgn}(\tau)}{|J_4
\tau|^{1/2}}, \ \ \ \ \ \ \ G(i \omega) =\pi^{1/4} \frac{i
\mathrm{sgn}(\omega)}{|J_4 \omega|^{1/2}}. \eeqn The diagram
Fig.~\ref{RG}$a$ leads to the following beta function for $u$:
\beqn \beta(u) = \frac{du}{d\ln l} =\frac{2}{\sqrt{\pi}}
\frac{1}{J_4} \sum_{i, j} |C_{ij}|^2 u^2=
\frac{2J^2}{\sqrt{\pi}J_4} u^2. \label{beta} \eeqn Here we have
replaced $\sum_{i, j} |C_{ij}|^2$ by $J^2$, which is consistent
with the distribution of $C_{ij}$, in the large $N$ limit.

Diagrams Fig.~\ref{RG}$b$ and $c$ will contribute at the
subleading order of $1/N$. For example, Fig.~\ref{RG}$b$ will
generate a term $ \sim \sum_{m,n} C_{im} C_{mn} C_{nj} C_{kl} u^2
\chi_i \chi_j \chi_k \chi_l$. This term is subleading in $1/N$
counting after disorder average.

The beta function indicates that the $H^\prime$ perturbation with
$u > 0$ ($u < 0$) is marginally relevant (marginally irrelevant)
at the SYK$_4$ fixed point. If we start with a small perturbation
$u > 0$, the RG equation implies that it will become order 1 at
the energy scale $\tilde{\Lambda}$ where \beqn \tilde{\Lambda}
\sim \Lambda \exp\left(- \frac{\sqrt{\pi}J_4}{2 J^2 u} \right).
\label{scale} \eeqn $\Lambda$ is the UV cut-off of the RG that we
can roughly take as $\Lambda \sim J_4$. The standard scaling
relation between the energy scale (mass gap) and the tuning
parameter $r$ away from a critical point $r_c$ is $\tilde{\Lambda}
\sim |r-r_c|^{\nu}$, thus the quantum phase transition led by
tuning $u$ across zero has exponent $\nu = + \infty$, which is
analogous to the Kosterlitz-Thouless transition~\cite{KT}.

This RG analysis predicts that the SYK model, although describes a
non-Fermi liquid state, actually has similar instabilities as the
ordinary Fermi liquid: there exists symmetry allowed four fermion
terms that are marginally relevant/irreleavant depending on their
sign. When $u$ is marginally relevant, our mean field solution in
the next subsection (and the analysis of $H^\prime$ in the
previous subsection) suggests that the fate of the SYK model is
also similar to the ordinary Fermi liquid: the system develops
long range correlation $\langle \hat{b}(0) \
\hat{b}(\tau)\rangle$, where $\hat{b}$ is the fermion-bilinear
operator defined in the previous subsection. The physics here is
analogous to the condensation of Cooper pair of the ordinary Fermi
liquid theory.

The effective action of Eq.~\ref{cluster} after a
Hubbard-Stratonovich transformation reads \beqn \mathcal{S}_{eff}
&=& \int d\tau \frac{1}{2}\sum_i \chi_i\partial_\tau\chi_i +
\sum_{ijkl}\left\{\frac{J_{ijkl}}{4!}\chi_i\chi_j\chi_k\chi_l
+ \frac{u}{2} C_{ij}C_{kl}\chi_i\chi_j\chi_k\chi_l\right\}\\
&=&\int d\tau \left(\frac{1}{2}\chi_i\partial_\tau\chi_i
+\frac{u}{2}b^2 - iu C_{jk} b
\chi_j\chi_k\right)+\frac{J_{ijkl}}{4!}\chi_i\chi_j\chi_k\chi_l
\eeqn The Hubbard-Stratonovich field $b$ is a real field. Einstein
summation convention is assumed in all the equations. The indices
are summed from $1$ to $N$ with the constraint that different
indices cannot take the same value. Now we can perform disorder
average on $J_{ijkl}$ and $C_{jk}$ with the distribution
Eq.~\ref{norm}. Assuming everything is replica diagonal
(justification of this assumption will be given in section IV),
the disorder-averaged action is equivalent to the following form:
\beqn \label{replica} \mathcal{S}_{eff} &=&\int d\tau
\frac{1}{2}\chi_i\partial_\tau\chi_i +\frac{u}{2}b^2 -
u^2\frac{J^2}{N^2}\int d\tau_1 d\tau_2 \ (b(\tau_1) \
b(\tau_2))(\chi_j(\tau_1)\chi_j(\tau_2))^2 \nn &&-
\frac{J_4^2}{8N^3}\int d\tau_1d\tau_2 \
(\chi_i(\tau_1)\chi_i(\tau_2))^4. \eeqn This disorder-averaged
action has an explicit O($N$) symmetry, the fermion carries a
vector representation of the O($N$).

\begin{figure}[tbp]
\begin{center}
\includegraphics[width=300pt]{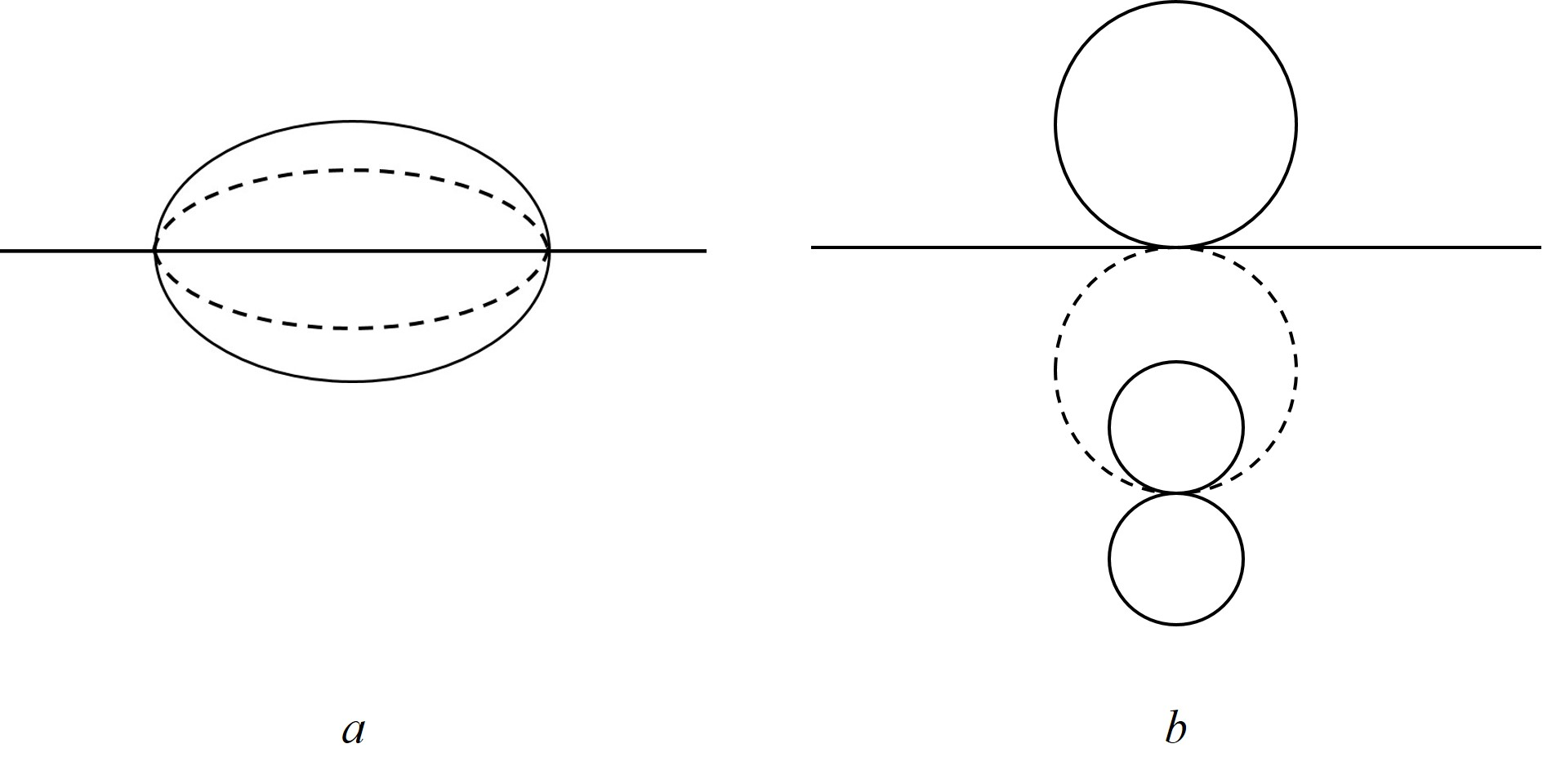}
\caption{ The fermion wave function renormalization based on
Eq.~\ref{cluster} and Eq.~\ref{replica} respectively. These
diagrams correspond to a $u^3$ term in the beta function, and it
carries a factor of $1/N$.} \label{bbx}
\end{center}
\end{figure}

The beta function for $u$ can also be computed based on
Eq.~\ref{replica}. Fig.~\ref{RG}$d$ based on Eq.~\ref{replica}
makes the same contribution to the beta function as
Fig.~\ref{RG}$a$. In the large$-N$ limit, the beta function
Eq.~\ref{beta} is actually exact. The higher order terms of the
beta function can be ignored in the large$-N$ limit even when $u$
grows beyond order-1 (and hence becomes dominant) under the RG
flow. For example the fermion wave function renormalization in
Fig.~\ref{bbx} corresponds to a $u^3$ term in the beta function,
and it carries a coefficient $1/N$. Other diagrams, such as the
ladder diagrams for the four-point functions computed in
Ref.~\onlinecite{MaldacenaStanford2016}, also contribute at the
subleading $1/N$ order compared with Fig.~\ref{RG}$a$,$d$.

\subsection{Mean field solution}

We can introduce fermion Green's function and Self-energy function
$G$ and $\Sigma$ by inserting the following integral in the action
($G$ and $\Sigma$ are real fields): \beq \int \mathcal{D}\Sigma
\mathcal{D} G
\exp\left\{-\frac{N}{2}\Sigma(\tau_1,\tau_2)\left(G(\tau_1,\tau_2)
-\frac{1}{N}\sum_i\chi_i(\tau_1)\chi_i(\tau_2)\right)\right\} \eeq
Then the action $S_{eff}$ is equivalent to: \beqn
\mathcal{S}_{eff} &=& - N \log \mathrm{Pf}\left( \partial_\tau -
\Sigma \right) + \int d\tau \ \frac{u}{2}b^2 - u^2J^2\int d\tau_1
d\tau_2 \ (b(\tau_1)b(\tau_2))(G(\tau_1,\tau_2))^2 \nn &&-
N\frac{J_4^2}{8}\int d\tau_1d\tau_2 \ (G(\tau_1,\tau_2))^4+N\int
d\tau_1 d\tau_2 \ \frac{1}{2}\Sigma(\tau_1,\tau_2)G(\tau_1,\tau_2)
\eeqn

Since the $H^\prime$ term itself has long range correlation of
$\hat{b}$, we expect that the phase with relevant $u$ perturbation
also develops the long range correlation of $b(\tau)$. Since the
ground state of $H^\prime$ has $b \sim N^{1/2}$, let us assume $
\langle b (\tau_1) b(\tau_2) \rangle = N w^2$, where $w$ takes
order-1 value with no time dependence. Then we can derive the mean
field equation for the Green's function, the self-energy, and also
$w$: \beq \label{green} G(i \omega_n)^{-1}= -i \omega_n -
\Sigma(i\omega_n) \eeq \beq \label{selfe} \Sigma(\tau) =
J_4^2G(\tau)^3 + 4u^2J^2w^2G(\tau) \eeq \beq \label{wsaddle} \int
d\tau\left(uJ^2G(\tau)^2 - \frac{1}{2}\delta(\tau)\right)uw = 0
\eeq

The saddle point Eq.~\ref{wsaddle} has two possible solutions:
$w=0$ or \beq \label{G2} \int d\tau \ G(\tau)^2 = \frac{1}{2uJ^2}.
\eeq For the $w=0$ saddle point, these equations return to the
saddle point equations for the pure $q = 4$ SYK model. The system
is in the chaotic non-Fermi liquid phase. However, when $w \neq
0$, in the low energy, the second term in Eq.~\ref{selfe} becomes
dominant, and the system is effectively described by a random two
fermion interaction and it is in a non-chaotic phase~\footnote{The
random four-fermion interaction, though irrelevant with the
presence of a random two-body interaction, still has perturbative
effect, and may lead to non-maximal chaos at finite temperature.
This effect was discussed in Ref.~\onlinecite{Banerjee2016}. Here
we still call this phase as non-chaotic phase, for conciseness.}.
In this phase, $G(\tau)$ will depend on the values of $w$, and we
can self-consistently determine $w$ from Eq.~\ref{G2}. The
chaotic-nonchaotic transition happens when $u$ is tuned from
negative to positive through 0. When $u$ is negative, Eq.~\ref{G2}
has no solution and $w$ has to be $0$. For any positive $u$, at
zero temperature there is always a solution with finite $w$. The
state with long range correlation $\langle b (0) b(\tau) \rangle$
spontaneously breaks the time-reversal symmetry $\mathcal{T}:
\chi_j \rightarrow \chi_j$.


There are two time scales in our problem, $\tau^{UV}_2\sim
(uwJ)^{-1}$ and $\tau^{UV}_4\sim J_4^{-1}$. In the small $u$
limit, namely $\tau_2^{UV}\gg\tau_4^{UV}$, the contribution of the
integral in Eq.~\ref{G2} mainly comes from the region $\tau
\in[\tau_4^{UV},\tau_2^{UV}]$, and in this region $G(\tau)$ takes
the form of the ordinary SYK model: \beq \int d\tau \
G(\tau)^2\simeq \int_{\tau_4^{UV}}^{\tau_2^{UV}} d\tau
\frac{2}{\sqrt{\pi}}\frac{1}{J_4\tau}=
\frac{2}{\sqrt{\pi}J_4}\log(\frac{J_4}{uwJ}) \eeq Together with
Eq.~\ref{G2}, we have \beq w \simeq \frac{J_4}{uJ}\exp \left( -
\frac{\sqrt{\pi}J_4}{4 uJ^2} \right). \label{wsolution}\eeq This
result is consistent with the observation that a positive $u$ is
only marginally relevant. The size of the condensate is analogous
to the superconductor gap of the BCS theory.

\begin{figure}[tbp]
\begin{center}
\includegraphics[width=250pt]{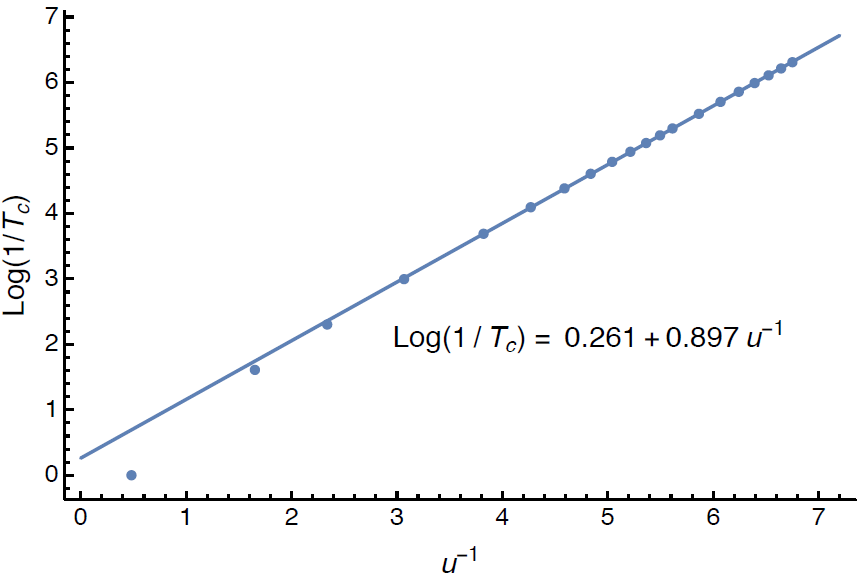}
\caption{Transition temperature $T_c$ as a function of $u$ by
numerically solving the mean field equations
(\ref{green}-\ref{wsaddle}). This confirms the scaling relation in
Eq.~\ref{scale}.} \label{MF}
\end{center}
\end{figure}

At finite $u$, the scale $\tilde{\Lambda}$ in Eq.~\ref{scale} can
be viewed as the critical temperature $T_c$ below which the system
develops nonzero $w$ and hence spontaneously breaks time-reversal
$\mathcal{T}$. Our numerical solution of the mean field equations
Eq.~\ref{green},\ref{selfe},\ref{wsaddle} confirms the scaling
between $T_c$ and $u$ (Fig.~\ref{MF}). In the numerical solution
we have taken $J^2/J_4 = 1$. Our RG Eq.~\ref{scale} predicts that
$T_c \sim \exp( - \frac{\sqrt{\pi}}{2} \frac{1}{u} ) = \exp(-
0.886/u)$, and our mean field solution gives $T_c \sim \exp(-
0.897/u)$.

\section{Further generalized perturbations}

Now let us consider a series of generalized Hamiltonians: \beqn
\label{cluster2} H = \mathrm{SYK}_q + H^\prime, \ \ \ \ H^\prime =
\frac{u}{2}\sum_{a=1}^M C^a_{ij}C^a_{kl}\chi_i\chi_j\chi_k\chi_l,
\eeqn with $M \sim N^A$. SYK$_q$ is the generalized SYK model with
a random $q-$fermion interaction, and $A \geq 0$. We first choose
the following normalization of $C^a_{ij}$ \beqn N^2 \
\overline{C^a_{ij}C^b_{kl}}=J^2 \delta_{ab}(\delta_{ik}\delta_{jl}
- \delta_{il}\delta_{jk}). \label{eq: C norm}\eeqn


We still start with the beta function of $u$. If we evaluate the
Green's functions at the SYK$_q$ fixed point, the beta function of
$u$ reads \beqn \beta(u) = \frac{du}{d\ln l} = (1 - \frac{4}{q}) u
+ C u^2 + \tilde{c}_3 \frac{M}{N} u^3 + \cdots \label{RGsyk}\eeqn
where $C > 0$ is an order-1 constant.

\subsection{cases with $A < 1$}

For $A < 1$, we can keep just the linear and quadratic terms of
the beta function, as all the higher order terms vanish in the
large$-N$ limit, when $u$ is order-1 or smaller. For $A < 1$ and
$u > 0$, $u$ is relevant at the SYK fixed point for $q > 4$, and
marginally relevant for $q=4$. We expect the system to behave
similarly as the case with $M = 1$ and $q =4$, namely the relevant
$u$ perturbation drives the system into a nonchaotic phase with
spontaneous $\mathcal{T}$ breaking: $\lim_{\tau \rightarrow
\infty} \sum_a \langle b^a(0) \ b^a(\tau) \rangle \neq 0$, where
$\hat{b}^a = i C^a_{jk}\chi_j\chi_k$. The same set of equations as
Eq.~\ref{green},\ref{selfe},\ref{wsaddle} can be derived, and in
this case $\sum_{a=1}^M \langle b^a(0) \ b^a(\tau) \rangle = N w^2
$, and $w$ is given by Eq.~\ref{wsolution}.

Exact diagonalization of the $H^\prime$ term in this case confirms
our expectations. To detect the long range correlation of $\langle
b^a(0) b^a(\tau)\rangle$, we measure the zero-frequency component
of the boson spectral function. The spectral function is defined
as \beqn D(\omega)=\frac{1}{M}\sum_{a=1}^M\sum_{n}\big|\langle
0|\hat{b}^a|n\rangle\big|^2\delta(\omega-E_n+E_0), \eeqn where
$E_n$ and $|n\rangle$ are eigenenergies and corresponding
eigenstates of the Hamiltonian $H'$, obtained from the exact
diagonalization $H'|n\rangle=E_n|n\rangle$ ($n=0,1,2,\cdots$).
$n=0$ labels the ground state. The $C_{ij}^a$ normalization in
Eq.~\ref{eq: C norm} ensures that
$\overline{\hat{b}^{a\dagger}\hat{b}^a}=1$ (the identity matrix)
in the large $N$ limit, so that $D(\omega)$ has a well-defined
thermodynamic limit. If the static correlation $D(\omega=0)$
remains finite in the thermodynamic limit $N\to\infty$, then the
system will develop long range correlation and spontaneously break
$\mathcal{T}$. The Fig.~\ref{fig: ED} shows the result of the
static correlation $D(\omega=0)$ (in logarithmic scale) for
different $N$ at $A=0.2$ and $u>0$. $\ln D(0)$ oscillates with $N$
in an eight-fold period due to the systematic change of
random-matrix ensemble of $H'$ as discussed in
Ref.~\onlinecite{you2016}. Apart from the oscillation,
$D(\omega=0)$ remains at and converges to a finite level (roughly
indicated by the dashed line in Fig.~\ref{fig: ED}). Therefore our
finite-sized calculation indeed supports a nonchaotic phase with
spontaneous $\mathcal{T}$ breaking for the $A<1$ and $u>0$ case.

\begin{figure}[htbp]
\begin{center}
\includegraphics[width=380pt]{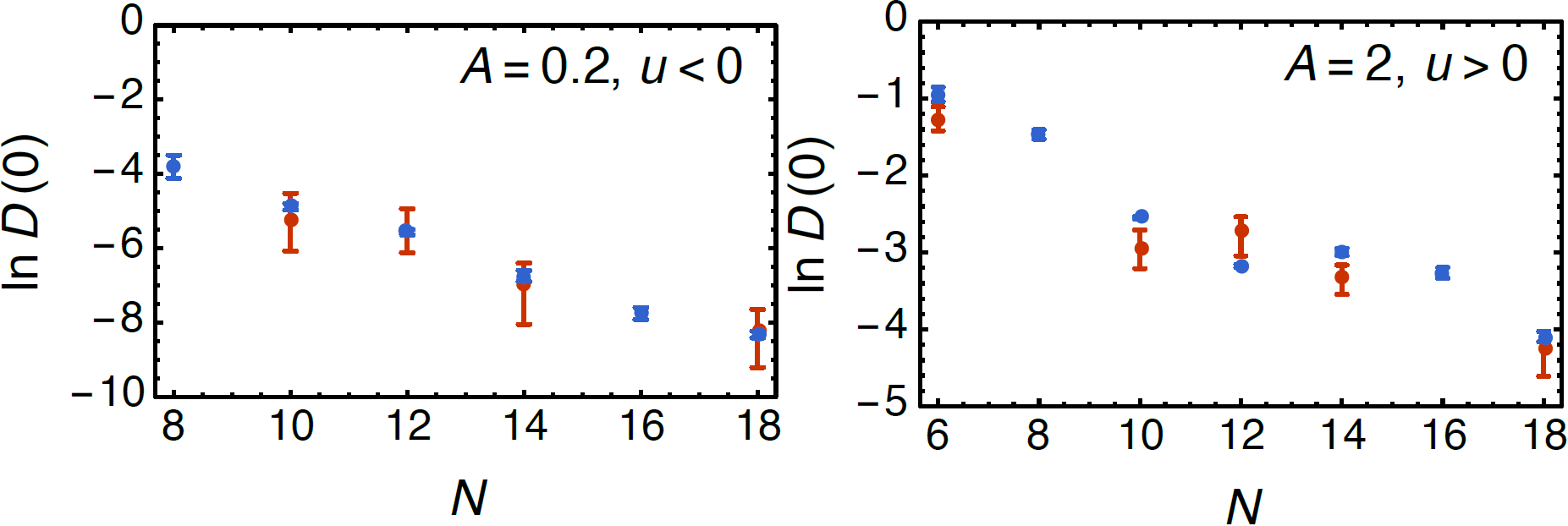}
\caption{The logarithmic static correlation $\ln D(0)$ v.s. the
fermion number $N$ for the case of $A = 0.2$, $u<0$ (left), and $A
=2$, $u >0$ (right). Neither case shows long range correlation of
the bosonic field $b^a$. Both $D(\omega = 0)$ (red) and
$\lim_{\omega\to 0}D(\omega)$ (blue) are plotted in the figures.}
\label{ED2}
\end{center}
\end{figure}

By contrast, for either $A > 1$, or $A < 1$ while $u < 0$, ED
shows $D(0)$ decreases rapidly with increasing $N$
(Fig.~\ref{ED2}).

\begin{figure}[htbp]
\begin{center}
\includegraphics[width=200pt]{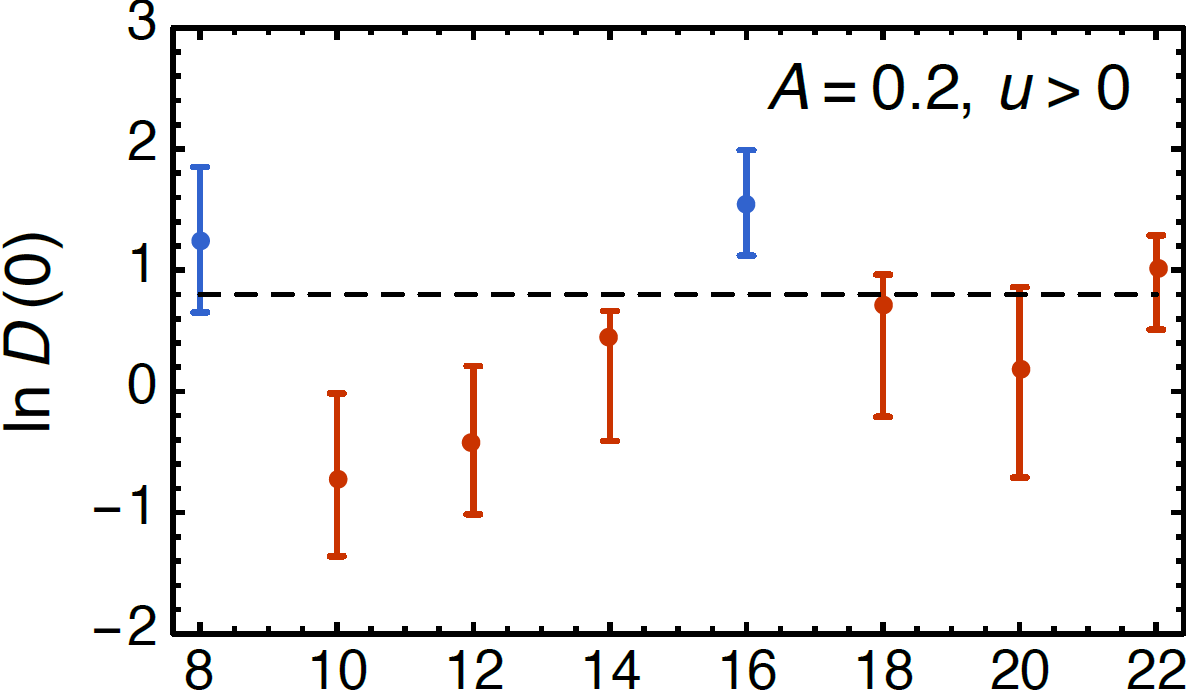}
\caption{The logarithmic static correlation $\ln D(0)$ v.s. the
fermion number $N$ for the case of $u > 0$ and $A = 0.2$. The
error bar shows the statistical deviation over different random
realizations of the coefficient $C_{ij}^a$. When $N\text{ mod
}8=0$, $D(\omega=0)$ vanishes exactly, so we use the finite
frequency extrapolation to obtain the static correlation
$D(0)=\lim_{\omega\to 0}D(\omega)$ in these cases.} \label{fig:
ED}
\end{center}
\end{figure}

For $A < 1$ and $u < 0$, the $u$ term flows to a stable fixed
point $u^\ast \sim - (1- 4/q)/C$. At this fixed point, since
$u^\ast$ is an order-1 number, the fermion self-energy correction
Fig.~\ref{bbx} is at the $M/N$ order, which vanishes in the
large$-N$ limit for $A < 1$. Thus the fermion scaling dimension
remains the same as the SYK$_q$ model: $\Delta_f = 1/q$. But at
this stable fixed point, the boson field $b^a \sim i
C^a_{jk}\chi_j \chi_k$ acquires a correction, and has scaling
dimension $\Delta_b = 1 - 2/q$ in the large$-N$ limit. Starting
with a SYK$_q$ model with $q > 4$, changing the sign of $u$ will
drive a chaotic-nonchaotic transition with exponent $\nu = q/(q -
4)$.


\subsection{cases with $A > 1$}

For $A > 1$, the RG equation is uncontrolled because the higher
order terms in the beta function dominate in the large$-N$ limit.
However, we can understand the model by taking the limit $M
\rightarrow + \infty$ first. One intuitive way to think about this
case is that according to the central limit theorem $\sum_{a =
1}^M C^a_{ij}C^a_{kl}$ with $M \rightarrow + \infty$ follows the
Gaussian distribution. So for either sign of $u$,
Eq.~\ref{cluster2} should behave the same as the $q = 4$ SYK
model. In order to explicitly demonstrate this statement, it is
more convenient to use a different normalization of $C^a_{ij}$:
\beq N^{(3+A) / 2} \ \overline{C^a_{ij}C^b_{kl}}=J^2
\delta_{ab}(\delta_{ik}\delta_{jl} - \delta_{il}\delta_{jk}).
\label{newnorm}\eeq  We can perform the disorder average and
integrating out $C^a_{ij}$, the leading order term in the
large$-N$ limit is an eight-fermion interaction term $\sim
\frac{u^2J^4}{N^3}\int\int d\tau
d\tau'(\chi_i(\tau)\chi_i(\tau'))^4$, just like the disorder
averaged $q=4$ SYK model, while all higher order $8n$-fermion
interaction terms $\mathcal{S}^{(8n)}$ are suppressed $\sim
\frac{(u^2J^4)^n}{N^{3n+A(n-1)}}\left(\int\int d\tau
d\tau'(\chi_i(\tau)\chi_i(\tau'))^4\right)^n$. Thus for $A
> 1$, the $u-$term actually behaves the same as the SYK model in
the large$-N$ limit. This conclusion is consistent with the
previous study of a similar generalization of the SYK
model~\cite{danshita}.

\subsection{the $H'$ term with $A = 1$}

$A = 1$ is the critical situation, and the $H^\prime$ term itself
(equivalent to taking $q = +\infty$ in Eq.~\ref{cluster2}) is
already interesting enough when $A = 1$. With the $H^\prime$ term
only, we numerically solve the following coupled Schwinger-Dyson
equations with the normalization from Eq.~\ref{newnorm}: \beq
\label{fgreen} \tilde{G}_f(i \omega_n)^{-1}= -i \omega_n -
\tilde{\Sigma}_f(i\omega_n), \ \ \ \Sigma_f(\tau)
=4\sqrt{\frac{M}{N}}u^2J^2G_b(\tau)G_f(\tau) \eeq \beq
\label{bgreen}
\tilde{G}_b(i\omega_n)^{-1}=u-\tilde{\Sigma}_b(i\omega_n), \ \ \
\Sigma_b(\tau)=2\sqrt{\frac{N}{M}}u^2J^2G_f^2(\tau) \eeq

\begin{figure}[tbp]
\begin{center}
\includegraphics[width=380pt]{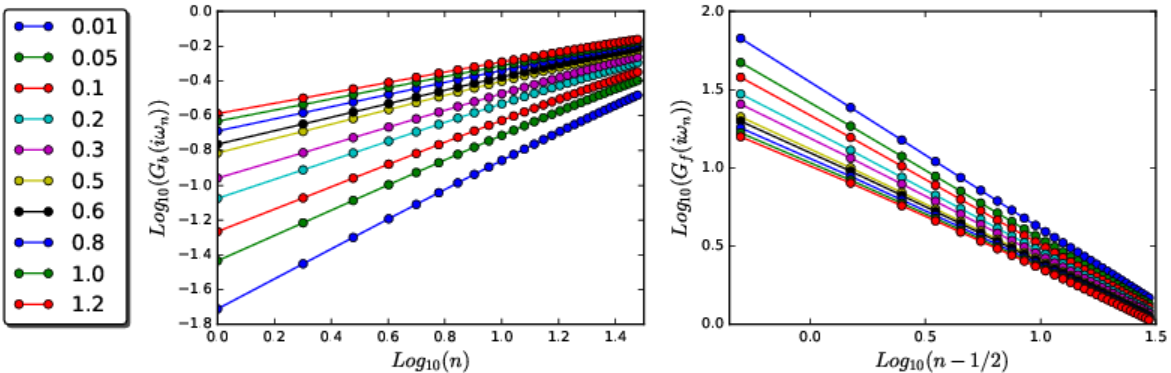}
\caption{The numerical solution of Eq.~\ref{fgreen},\ref{bgreen},
for $u = -1$, $J = 1$, $\beta = 300$ with different $M/N$, without
assuming a conformal solution from the beginning. Both the boson
and fermion Green's functions have nice power-law scaling with the
frequency, whose scaling dimensions depend on $M/N$.}
\label{scaling}
\end{center}
\end{figure}

\begin{figure}[tbp]
\begin{center}
\includegraphics[width=250pt]{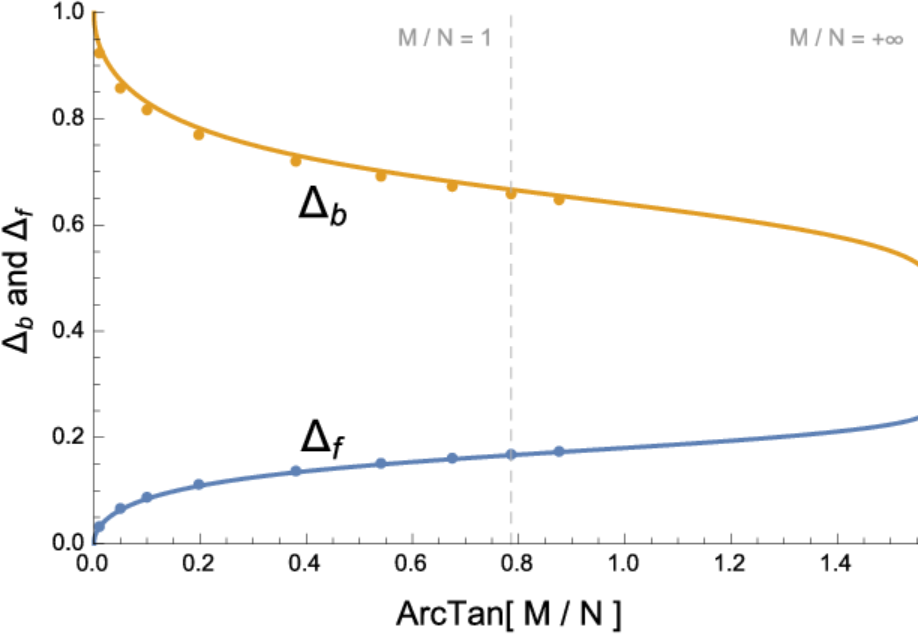}
\caption{We numerically solve the Schwinger-Dyson equations
(\ref{fgreen}-\ref{bgreen}) for $u=-1, J=1, \beta=300$ and fit the
low frequency part as a power law. The scaling dimensions are
continuous function of $M/N$, and for all the data points, the
relation $2\Delta_f+\Delta_b=1$ is held. The solid curves plot the
solution of the scaling dimensions based on Eq.~\ref{deltab}. In
particular, for $M/N=1$ (the dashed line), the scaling dimensions
obtained from both the numerical and analytical solutions match
with the prediction from the SUSY SYK model. \cite{FuSUSY2016}}
\label{scalingdim}
\end{center}
\end{figure}

For the case with $A = 1$ and $u < 0$, the numerical solution of
Eq.~\ref{fgreen},\ref{bgreen} generates well-converged power-law
correlation functions for all $\alpha = M/N$, for both the fermion
and boson fields (Fig.~\ref{scaling}). And the scaling dimensions
always satisfy $2\Delta_f + \Delta_b = 1$.

Alternatively, by assuming that $ G_b (\tau) \sim
B/|\tau|^{2\Delta_b} $ and $ G_f (\tau) \sim F \
\mathrm{sgn}(\tau) /|\tau|^{2\Delta_f} $ in the infrared limit,
Eq.~\ref{fgreen},\ref{bgreen} reduce to the following equation for
$\Delta_b$ for each ratio $M/N$: \beqn 2 \frac{M}{N}
\frac{\sin^2\left( \frac{\pi}{2} \Delta_b \right)}{\sin^2 \left(
\pi \Delta_b \right)} \frac{\Gamma(\Delta_b)\Gamma(- \Delta_b)}{
\Gamma(2\Delta_b) \Gamma(-2\Delta_b)} \frac{2\Delta_b -
1}{2\Delta_b} = -1. \label{deltab}\eeqn $\Delta_f $ can be
determined by $\Delta_b + 2 \Delta_f = 1$. In particular, for $M/N
= 1$, our solution matches with the result of the SUSY SYK
model~\cite{FuSUSY2016}, where the model also has $M/N = 1$ and $u
< 0$. The numerical solutions of Eq.~\ref{fgreen},\ref{bgreen} and
analytical solution of Eq.~\ref{deltab} are both plotted in
Fig.~\ref{scalingdim}. With small $M/N$, $\Delta_f$ is
approximately $\Delta_f \sim 1/\pi \sqrt{M/N}$.

\section{Summary and Discussion}

In this work we have demonstrated through various methods that the
non-Fermi liquid fixed point of the SYK$_4$ model is instable
against a class of marginally relevant four fermion perturbations,
and these perturbations drive the system into a non-chaotic state
with zero ground state entropy, and spontaneous time-reversal
symmetry breaking. Because these perturbations are only marginally
relevant, this effect occurs at exponentially low energy scale for
a fixed strength of the perturbation. Spontaneous time-reversal
symmetry breaking in experimental systems can be probed through
Kerr rotation, which has been successfully applied to various
condensed matter systems~\cite{Kerr1,Kerr2,Kerr3,Kerr4}. Similar
perturbations (with an opposite sign) can drive the SYK$_q$ model
with $q > 4$ to a series of fixed points with continuously varying
scaling dimensions.

So far we have ignored the replica index, for instance in
Eq.~\ref{replica}. We will provide a self-consistent justification
for this procedure. The usual argument for ignoring the replica
index after disorder averaging the SYK interaction $J_{ijkl}$ is
that, the replica off-diagonal terms are subleading in $1/N$
expansion~\cite{Gu2016}. Here we will investigate the replica
index introduced after disorder averaging $C^a_{jk}$, and we only
need to consider the case with $A \leq 1$, since as we have argued
before, the case with $A > 1$ is equivalent to the SYK$_4$ model.

Starting with the boson-fermion interaction term, $- iu C^a_{jk}
b_a \chi_j\chi_k$, reinstating the replica index after
disorder-average will lead to the following term \beqn
\label{offd} \sim - \frac{u^2 J^2}{N^2} \sum_{\alpha,\beta} \int
d\tau \int d\tau^\prime \sum_{a = 1}^M
b^a_\alpha(\tau)b^a_\beta(\tau^\prime) \left( \chi^\alpha_j(\tau)
\chi^\beta_j(\tau^\prime)\right)^2. \eeqn In the phase where
$b^a_\alpha$ does not condense (corresponds to $u < 0$ in our
case), the usual perturbation argument like
Ref.~\onlinecite{Gu2016} will conclude that the replica
off-diagonal terms will always make subleading contribution to the
partition function compared with the diagonal terms. In the phase
with $b^a$ condenses ($A < 1$, $u>0$), the mean field solution
tells us that $\sum_{a = 1}^M \langle
b^a_\alpha(\tau)b^a_\beta(\tau^\prime) \rangle $ in Eq.~\ref{offd}
is at order of $N$. Then the perturbation argument will tell us
when $u > 0$ and $A < 1$, the contribution from the replica
off-diagonal terms is still subleading. Thus for all the main
conclusions of this work, we can always make the replica diagonal
assumption, and hence ignore the replica index.

The authors thank Wenbo Fu, Yingfei Gu, Xiao-Liang Qi, Subir
Sachdev for very helpful discussions. Zhen Bi and Cenke Xu are
supported by the David and Lucile Packard Foundation and NSF Grant
No. DMR-1151208.

\bibliography{SYK}

\begin{thebibliography}{42}
\expandafter\ifx\csname natexlab\endcsname\relax\def\natexlab#1{#1}\fi
\expandafter\ifx\csname bibnamefont\endcsname\relax
  \def\bibnamefont#1{#1}\fi
\expandafter\ifx\csname bibfnamefont\endcsname\relax
  \def\bibfnamefont#1{#1}\fi
\expandafter\ifx\csname citenamefont\endcsname\relax
  \def\citenamefont#1{#1}\fi
\expandafter\ifx\csname url\endcsname\relax
  \def\url#1{\texttt{#1}}\fi
\expandafter\ifx\csname urlprefix\endcsname\relax\def\urlprefix{URL }\fi
\providecommand{\bibinfo}[2]{#2}
\providecommand{\eprint}[2][]{\url{#2}}

\bibitem[{\citenamefont{Hertz}(1976)}]{hertz}
\bibinfo{author}{\bibfnamefont{J.~A.} \bibnamefont{Hertz}},
  \bibinfo{journal}{Phys. Rev. B} \textbf{\bibinfo{volume}{14}},
  \bibinfo{pages}{1165} (\bibinfo{year}{1976}),
  \urlprefix\url{http://link.aps.org/doi/10.1103/PhysRevB.14.1165}.

\bibitem[{\citenamefont{Millis}(1993)}]{millis}
\bibinfo{author}{\bibfnamefont{A.~J.} \bibnamefont{Millis}},
  \bibinfo{journal}{Phys. Rev. B} \textbf{\bibinfo{volume}{48}},
  \bibinfo{pages}{7183} (\bibinfo{year}{1993}),
  \urlprefix\url{http://link.aps.org/doi/10.1103/PhysRevB.48.7183}.

\bibitem[{\citenamefont{L\"ohneysen et~al.}(2007)\citenamefont{L\"ohneysen,
  Rosch, Vojta, and W\"olfle}}]{nfl}
\bibinfo{author}{\bibfnamefont{H.~v.} \bibnamefont{L\"ohneysen}},
  \bibinfo{author}{\bibfnamefont{A.}~\bibnamefont{Rosch}},
  \bibinfo{author}{\bibfnamefont{M.}~\bibnamefont{Vojta}}, \bibnamefont{and}
  \bibinfo{author}{\bibfnamefont{P.}~\bibnamefont{W\"olfle}},
  \bibinfo{journal}{Rev. Mod. Phys.} \textbf{\bibinfo{volume}{79}},
  \bibinfo{pages}{1015} (\bibinfo{year}{2007}),
  \urlprefix\url{http://link.aps.org/doi/10.1103/RevModPhys.79.1015}.

\bibitem[{\citenamefont{Lee}(2009)}]{lee2009}
\bibinfo{author}{\bibfnamefont{S.-S.} \bibnamefont{Lee}},
  \bibinfo{journal}{Phys. Rev. B} \textbf{\bibinfo{volume}{80}},
  \bibinfo{pages}{165102} (\bibinfo{year}{2009}),
  \urlprefix\url{http://link.aps.org/doi/10.1103/PhysRevB.80.165102}.

\bibitem[{\citenamefont{Mross et~al.}(2010)\citenamefont{Mross, McGreevy, Liu,
  and Senthil}}]{senthilnfl}
\bibinfo{author}{\bibfnamefont{D.~F.} \bibnamefont{Mross}},
  \bibinfo{author}{\bibfnamefont{J.}~\bibnamefont{McGreevy}},
  \bibinfo{author}{\bibfnamefont{H.}~\bibnamefont{Liu}}, \bibnamefont{and}
  \bibinfo{author}{\bibfnamefont{T.}~\bibnamefont{Senthil}},
  \bibinfo{journal}{Phys. Rev. B} \textbf{\bibinfo{volume}{82}},
  \bibinfo{pages}{045121} (\bibinfo{year}{2010}),
  \urlprefix\url{http://link.aps.org/doi/10.1103/PhysRevB.82.045121}.

\bibitem[{\citenamefont{Metlitski and Sachdev}(2010{\natexlab{a}})}]{maxnfl1}
\bibinfo{author}{\bibfnamefont{M.~A.} \bibnamefont{Metlitski}}
  \bibnamefont{and} \bibinfo{author}{\bibfnamefont{S.}~\bibnamefont{Sachdev}},
  \bibinfo{journal}{Phys. Rev. B} \textbf{\bibinfo{volume}{82}},
  \bibinfo{pages}{075127} (\bibinfo{year}{2010}{\natexlab{a}}),
  \urlprefix\url{http://link.aps.org/doi/10.1103/PhysRevB.82.075127}.

\bibitem[{\citenamefont{Metlitski and Sachdev}(2010{\natexlab{b}})}]{maxnfl2}
\bibinfo{author}{\bibfnamefont{M.~A.} \bibnamefont{Metlitski}}
  \bibnamefont{and} \bibinfo{author}{\bibfnamefont{S.}~\bibnamefont{Sachdev}},
  \bibinfo{journal}{Phys. Rev. B} \textbf{\bibinfo{volume}{82}},
  \bibinfo{pages}{075128} (\bibinfo{year}{2010}{\natexlab{b}}),
  \urlprefix\url{http://link.aps.org/doi/10.1103/PhysRevB.82.075128}.

\bibitem[{\citenamefont{Dalidovich and Lee}(2013)}]{leenfl}
\bibinfo{author}{\bibfnamefont{D.}~\bibnamefont{Dalidovich}} \bibnamefont{and}
  \bibinfo{author}{\bibfnamefont{S.-S.} \bibnamefont{Lee}},
  \bibinfo{journal}{Phys. Rev. B} \textbf{\bibinfo{volume}{88}},
  \bibinfo{pages}{245106} (\bibinfo{year}{2013}),
  \urlprefix\url{http://link.aps.org/doi/10.1103/PhysRevB.88.245106}.

\bibitem[{\citenamefont{{Sachdev} and {Ye}}(1993)}]{SachdevYe1993}
\bibinfo{author}{\bibfnamefont{S.}~\bibnamefont{{Sachdev}}} \bibnamefont{and}
  \bibinfo{author}{\bibfnamefont{J.}~\bibnamefont{{Ye}}},
  \bibinfo{journal}{Physical Review Letters} \textbf{\bibinfo{volume}{70}},
  \bibinfo{pages}{3339} (\bibinfo{year}{1993}), \eprint{cond-mat/9212030}.

\bibitem[{\citenamefont{{Kitaev}}(2015)}]{Kitaev2015}
\bibinfo{author}{\bibfnamefont{A.}~\bibnamefont{{Kitaev}}},
  \emph{\bibinfo{title}{{A simple model of quantum holography}}},
  \bibinfo{howpublished}{\url{http://online.kitp.ucsb.edu/online/entangled15/k%
itaev/,http: //online.kitp.ucsb.edu/online/entangled15/kitaev2/.}}
  (\bibinfo{year}{2015}), \bibinfo{note}{{T}alks at KITP, April 7, 2015 and May
  27, 2015.}

\bibitem[{\citenamefont{{Maldacena} and
  {Stanford}}(2016)}]{MaldacenaStanford2016}
\bibinfo{author}{\bibfnamefont{J.}~\bibnamefont{{Maldacena}}} \bibnamefont{and}
  \bibinfo{author}{\bibfnamefont{D.}~\bibnamefont{{Stanford}}},
  \bibinfo{journal}{\prd} \textbf{\bibinfo{volume}{94}}, \bibinfo{eid}{106002}
  (\bibinfo{year}{2016}), \eprint{1604.07818}.

\bibitem[{\citenamefont{{Maldacena}
  et~al.}(2016{\natexlab{a}})\citenamefont{{Maldacena}, {Shenker}, and
  {Stanford}}}]{MSSbound}
\bibinfo{author}{\bibfnamefont{J.}~\bibnamefont{{Maldacena}}},
  \bibinfo{author}{\bibfnamefont{S.~H.} \bibnamefont{{Shenker}}},
  \bibnamefont{and}
  \bibinfo{author}{\bibfnamefont{D.}~\bibnamefont{{Stanford}}},
  \bibinfo{journal}{Journal of High Energy Physics}
  \textbf{\bibinfo{volume}{8}}, \bibinfo{eid}{106}
  (\bibinfo{year}{2016}{\natexlab{a}}), \eprint{1503.01409}.

\bibitem[{\citenamefont{Sachdev}(2010)}]{Sachdev2010}
\bibinfo{author}{\bibfnamefont{S.}~\bibnamefont{Sachdev}},
  \bibinfo{journal}{Phys. Rev. Lett.} \textbf{\bibinfo{volume}{105}},
  \bibinfo{pages}{151602} (\bibinfo{year}{2010}),
  \urlprefix\url{http://link.aps.org/doi/10.1103/PhysRevLett.105.151602}.

\bibitem[{\citenamefont{{Sachdev}}(2015)}]{Sachdev2015}
\bibinfo{author}{\bibfnamefont{S.}~\bibnamefont{{Sachdev}}},
  \bibinfo{journal}{Physical Review X} \textbf{\bibinfo{volume}{5}},
  \bibinfo{eid}{041025} (\bibinfo{year}{2015}), \eprint{1506.05111}.

\bibitem[{\citenamefont{{Polchinski} and {Rosenhaus}}(2016)}]{Polchinski2016}
\bibinfo{author}{\bibfnamefont{J.}~\bibnamefont{{Polchinski}}}
  \bibnamefont{and}
  \bibinfo{author}{\bibfnamefont{V.}~\bibnamefont{{Rosenhaus}}},
  \bibinfo{journal}{Journal of High Energy Physics}
  \textbf{\bibinfo{volume}{4}}, \bibinfo{eid}{1} (\bibinfo{year}{2016}),
  \eprint{1601.06768}.

\bibitem[{\citenamefont{{Maldacena}
  et~al.}(2016{\natexlab{b}})\citenamefont{{Maldacena}, {Stanford}, and
  {Yang}}}]{MSY2016}
\bibinfo{author}{\bibfnamefont{J.}~\bibnamefont{{Maldacena}}},
  \bibinfo{author}{\bibfnamefont{D.}~\bibnamefont{{Stanford}}},
  \bibnamefont{and} \bibinfo{author}{\bibfnamefont{Z.}~\bibnamefont{{Yang}}},
  \bibinfo{journal}{ArXiv e-prints}  (\bibinfo{year}{2016}{\natexlab{b}}),
  \eprint{1606.01857}.

\bibitem[{\citenamefont{Jensen}(2016)}]{jensen2016}
\bibinfo{author}{\bibfnamefont{K.}~\bibnamefont{Jensen}},
  \bibinfo{journal}{Phys. Rev. Lett.} \textbf{\bibinfo{volume}{117}},
  \bibinfo{pages}{111601} (\bibinfo{year}{2016}),
  \urlprefix\url{http://link.aps.org/doi/10.1103/PhysRevLett.117.111601}.

\bibitem[{\citenamefont{Engels{\"o}y et~al.}(2016)\citenamefont{Engels{\"o}y,
  Mertens, and Verlinde}}]{verlinde2016}
\bibinfo{author}{\bibfnamefont{J.}~\bibnamefont{Engels{\"o}y}},
  \bibinfo{author}{\bibfnamefont{T.~G.} \bibnamefont{Mertens}},
  \bibnamefont{and} \bibinfo{author}{\bibfnamefont{H.}~\bibnamefont{Verlinde}},
  \bibinfo{journal}{Journal of High Energy Physics}
  \textbf{\bibinfo{volume}{2016}}, \bibinfo{pages}{139} (\bibinfo{year}{2016}),
  ISSN \bibinfo{issn}{1029-8479},
  \urlprefix\url{http://dx.doi.org/10.1007/JHEP07(2016)139}.

\bibitem[{\citenamefont{{Georges} et~al.}(2001)\citenamefont{{Georges},
  {Parcollet}, and {Sachdev}}}]{Georges2001}
\bibinfo{author}{\bibfnamefont{A.}~\bibnamefont{{Georges}}},
  \bibinfo{author}{\bibfnamefont{O.}~\bibnamefont{{Parcollet}}},
  \bibnamefont{and}
  \bibinfo{author}{\bibfnamefont{S.}~\bibnamefont{{Sachdev}}},
  \bibinfo{journal}{\prb} \textbf{\bibinfo{volume}{63}}, \bibinfo{eid}{134406}
  (\bibinfo{year}{2001}), \eprint{cond-mat/0009388}.

\bibitem[{\citenamefont{You et~al.}(2016)\citenamefont{You, Ludwig, and
  Xu}}]{you2016}
\bibinfo{author}{\bibfnamefont{Y.~Z.} \bibnamefont{You}},
  \bibinfo{author}{\bibfnamefont{A.~W.~W.} \bibnamefont{Ludwig}},
  \bibnamefont{and} \bibinfo{author}{\bibfnamefont{C.}~\bibnamefont{Xu}},
  \bibinfo{journal}{arXiv:1602.06964}  (\bibinfo{year}{2016}).

\bibitem[{\citenamefont{{Fu} and {Sachdev}}(2016)}]{Fu2016}
\bibinfo{author}{\bibfnamefont{W.}~\bibnamefont{{Fu}}} \bibnamefont{and}
  \bibinfo{author}{\bibfnamefont{S.}~\bibnamefont{{Sachdev}}},
  \bibinfo{journal}{\prb} \textbf{\bibinfo{volume}{94}}, \bibinfo{eid}{035135}
  (\bibinfo{year}{2016}), \eprint{1603.05246}.

\bibitem[{\citenamefont{{Cotler} et~al.}(2016)\citenamefont{{Cotler},
  {Gur-Ari}, {Hanada}, {Polchinski}, {Saad}, {Shenker}, {Stanford},
  {Streicher}, and {Tezuka}}}]{Cotler2016}
\bibinfo{author}{\bibfnamefont{J.~S.} \bibnamefont{{Cotler}}},
  \bibinfo{author}{\bibfnamefont{G.}~\bibnamefont{{Gur-Ari}}},
  \bibinfo{author}{\bibfnamefont{M.}~\bibnamefont{{Hanada}}},
  \bibinfo{author}{\bibfnamefont{J.}~\bibnamefont{{Polchinski}}},
  \bibinfo{author}{\bibfnamefont{P.}~\bibnamefont{{Saad}}},
  \bibinfo{author}{\bibfnamefont{S.~H.} \bibnamefont{{Shenker}}},
  \bibinfo{author}{\bibfnamefont{D.}~\bibnamefont{{Stanford}}},
  \bibinfo{author}{\bibfnamefont{A.}~\bibnamefont{{Streicher}}},
  \bibnamefont{and} \bibinfo{author}{\bibfnamefont{M.}~\bibnamefont{{Tezuka}}},
  \bibinfo{journal}{ArXiv e-prints}  (\bibinfo{year}{2016}),
  \eprint{1611.04650}.

\bibitem[{\citenamefont{{Fu} et~al.}(2016)\citenamefont{{Fu}, {Gaiotto},
  {Maldacena}, and {Sachdev}}}]{FuSUSY2016}
\bibinfo{author}{\bibfnamefont{W.}~\bibnamefont{{Fu}}},
  \bibinfo{author}{\bibfnamefont{D.}~\bibnamefont{{Gaiotto}}},
  \bibinfo{author}{\bibfnamefont{J.}~\bibnamefont{{Maldacena}}},
  \bibnamefont{and}
  \bibinfo{author}{\bibfnamefont{S.}~\bibnamefont{{Sachdev}}},
  \bibinfo{journal}{ArXiv e-prints}  (\bibinfo{year}{2016}),
  \eprint{1610.08917}.

\bibitem[{\citenamefont{{Gu} et~al.}(2016)\citenamefont{{Gu}, {Qi}, and
  {Stanford}}}]{Gu2016}
\bibinfo{author}{\bibfnamefont{Y.}~\bibnamefont{{Gu}}},
  \bibinfo{author}{\bibfnamefont{X.-L.} \bibnamefont{{Qi}}}, \bibnamefont{and}
  \bibinfo{author}{\bibfnamefont{D.}~\bibnamefont{{Stanford}}},
  \bibinfo{journal}{ArXiv e-prints}  (\bibinfo{year}{2016}),
  \eprint{1609.07832}.

\bibitem[{\citenamefont{{Davison} et~al.}(2016)\citenamefont{{Davison}, {Fu},
  {Georges}, {Gu}, {Jensen}, and {Sachdev}}}]{Davison2016}
\bibinfo{author}{\bibfnamefont{R.~A.} \bibnamefont{{Davison}}},
  \bibinfo{author}{\bibfnamefont{W.}~\bibnamefont{{Fu}}},
  \bibinfo{author}{\bibfnamefont{A.}~\bibnamefont{{Georges}}},
  \bibinfo{author}{\bibfnamefont{Y.}~\bibnamefont{{Gu}}},
  \bibinfo{author}{\bibfnamefont{K.}~\bibnamefont{{Jensen}}}, \bibnamefont{and}
  \bibinfo{author}{\bibfnamefont{S.}~\bibnamefont{{Sachdev}}},
  \bibinfo{journal}{ArXiv e-prints}  (\bibinfo{year}{2016}),
  \eprint{1612.00849}.

\bibitem[{\citenamefont{{Witten}}(2016)}]{Witten2016}
\bibinfo{author}{\bibfnamefont{E.}~\bibnamefont{{Witten}}},
  \bibinfo{journal}{ArXiv e-prints}  (\bibinfo{year}{2016}),
  \eprint{1610.09758}.

\bibitem[{\citenamefont{{Klebanov} and {Tarnopolsky}}(2016)}]{Klebanov2016}
\bibinfo{author}{\bibfnamefont{I.~R.} \bibnamefont{{Klebanov}}}
  \bibnamefont{and}
  \bibinfo{author}{\bibfnamefont{G.}~\bibnamefont{{Tarnopolsky}}},
  \bibinfo{journal}{ArXiv e-prints}  (\bibinfo{year}{2016}),
  \eprint{1611.08915}.

\bibitem[{\citenamefont{{Turiaci} and {Verlinde}}(2017)}]{Verlinde2017}
\bibinfo{author}{\bibfnamefont{G.}~\bibnamefont{{Turiaci}}} \bibnamefont{and}
  \bibinfo{author}{\bibfnamefont{H.}~\bibnamefont{{Verlinde}}},
  \bibinfo{journal}{ArXiv e-prints}  (\bibinfo{year}{2017}),
  \eprint{1701.00528}.

\bibitem[{\citenamefont{{Banerjee} and {Altman}}(2016)}]{Banerjee2016}
\bibinfo{author}{\bibfnamefont{S.}~\bibnamefont{{Banerjee}}} \bibnamefont{and}
  \bibinfo{author}{\bibfnamefont{E.}~\bibnamefont{{Altman}}},
  \bibinfo{journal}{ArXiv e-prints}  (\bibinfo{year}{2016}),
  \eprint{1610.04619}.

\bibitem[{\citenamefont{Gross and Rosenhaus}(2016)}]{gross2016}
\bibinfo{author}{\bibfnamefont{D.~J.} \bibnamefont{Gross}} \bibnamefont{and}
  \bibinfo{author}{\bibfnamefont{V.}~\bibnamefont{Rosenhaus}},
  \bibinfo{journal}{arXiv:1610.01569}  (\bibinfo{year}{2016}).

\bibitem[{\citenamefont{Garc\'{\i}a-Garc\'{\i}a and
  Verbaarschot}(2016)}]{antonio2016}
\bibinfo{author}{\bibfnamefont{A.~M.} \bibnamefont{Garc\'{\i}a-Garc\'{\i}a}}
  \bibnamefont{and} \bibinfo{author}{\bibfnamefont{J.~J.~M.}
  \bibnamefont{Verbaarschot}}, \bibinfo{journal}{Phys. Rev. D}
  \textbf{\bibinfo{volume}{94}}, \bibinfo{pages}{126010}
  (\bibinfo{year}{2016}),
  \urlprefix\url{http://link.aps.org/doi/10.1103/PhysRevD.94.126010}.

\bibitem[{\citenamefont{Krishnan et~al.}(2016)\citenamefont{Krishnan, Sanyal,
  and Subramanian}}]{krishnan2016}
\bibinfo{author}{\bibfnamefont{C.}~\bibnamefont{Krishnan}},
  \bibinfo{author}{\bibfnamefont{S.}~\bibnamefont{Sanyal}}, \bibnamefont{and}
  \bibinfo{author}{\bibfnamefont{P.~N.~B.} \bibnamefont{Subramanian}},
  \bibinfo{journal}{arXiv:1612.06330}  (\bibinfo{year}{2016}).

\bibitem[{\citenamefont{Metlitski et~al.}(2015)\citenamefont{Metlitski, Mross,
  Sachdev, and Senthil}}]{maxnfl}
\bibinfo{author}{\bibfnamefont{M.~A.} \bibnamefont{Metlitski}},
  \bibinfo{author}{\bibfnamefont{D.~F.} \bibnamefont{Mross}},
  \bibinfo{author}{\bibfnamefont{S.}~\bibnamefont{Sachdev}}, \bibnamefont{and}
  \bibinfo{author}{\bibfnamefont{T.}~\bibnamefont{Senthil}},
  \bibinfo{journal}{Phys. Rev. B} \textbf{\bibinfo{volume}{91}},
  \bibinfo{pages}{115111} (\bibinfo{year}{2015}),
  \urlprefix\url{http://link.aps.org/doi/10.1103/PhysRevB.91.115111}.

\bibitem[{\citenamefont{Kitaev}(2009)}]{kitaevclass}
\bibinfo{author}{\bibfnamefont{A.}~\bibnamefont{Kitaev}}, \bibinfo{journal}{AIP
  Conf. Proc} \textbf{\bibinfo{volume}{1134}}, \bibinfo{pages}{22}
  (\bibinfo{year}{2009}).

\bibitem[{\citenamefont{Schnyder et~al.}(2009)\citenamefont{Schnyder, Ryu,
  Furusaki, and Ludwig}}]{ludwigclass1}
\bibinfo{author}{\bibfnamefont{A.~P.} \bibnamefont{Schnyder}},
  \bibinfo{author}{\bibfnamefont{S.}~\bibnamefont{Ryu}},
  \bibinfo{author}{\bibfnamefont{A.}~\bibnamefont{Furusaki}}, \bibnamefont{and}
  \bibinfo{author}{\bibfnamefont{A.~W.~W.} \bibnamefont{Ludwig}},
  \bibinfo{journal}{AIP Conf. Proc.} \textbf{\bibinfo{volume}{1134}},
  \bibinfo{pages}{10} (\bibinfo{year}{2009}).

\bibitem[{\citenamefont{Ryu et~al.}(2010)\citenamefont{Ryu, Schnyder, Furusaki,
  and Ludwig}}]{ludwigclass2}
\bibinfo{author}{\bibfnamefont{S.}~\bibnamefont{Ryu}},
  \bibinfo{author}{\bibfnamefont{A.}~\bibnamefont{Schnyder}},
  \bibinfo{author}{\bibfnamefont{A.}~\bibnamefont{Furusaki}}, \bibnamefont{and}
  \bibinfo{author}{\bibfnamefont{A.}~\bibnamefont{Ludwig}},
  \bibinfo{journal}{New J. Phys.} \textbf{\bibinfo{volume}{12}},
  \bibinfo{pages}{065010} (\bibinfo{year}{2010}).

\bibitem[{\citenamefont{Kosterlit and Thouless}(1973)}]{KT}
\bibinfo{author}{\bibfnamefont{J.~M.} \bibnamefont{Kosterlit}}
  \bibnamefont{and} \bibinfo{author}{\bibfnamefont{D.~J.}
  \bibnamefont{Thouless}}, \bibinfo{journal}{J. Phys. C: Solid State Phys}
  \textbf{\bibinfo{volume}{6}}, \bibinfo{pages}{1181} (\bibinfo{year}{1973}).

\bibitem[{\citenamefont{Danshita et~al.}(2016)\citenamefont{Danshita, Hanada,
  and Tezuka}}]{danshita}
\bibinfo{author}{\bibfnamefont{I.}~\bibnamefont{Danshita}},
  \bibinfo{author}{\bibfnamefont{M.}~\bibnamefont{Hanada}}, \bibnamefont{and}
  \bibinfo{author}{\bibfnamefont{M.}~\bibnamefont{Tezuka}},
  \bibinfo{journal}{arXiv:1606.02454}  (\bibinfo{year}{2016}).

\bibitem[{\citenamefont{Kapitulnik et~al.}(2009)\citenamefont{Kapitulnik, Xia,
  Schemm, and Palevski}}]{Kerr1}
\bibinfo{author}{\bibfnamefont{A.}~\bibnamefont{Kapitulnik}},
  \bibinfo{author}{\bibfnamefont{J.}~\bibnamefont{Xia}},
  \bibinfo{author}{\bibfnamefont{E.}~\bibnamefont{Schemm}}, \bibnamefont{and}
  \bibinfo{author}{\bibfnamefont{A.}~\bibnamefont{Palevski}},
  \bibinfo{journal}{New Journal of Physics} \textbf{\bibinfo{volume}{11}},
  \bibinfo{pages}{055060} (\bibinfo{year}{2009}),
  \urlprefix\url{http://stacks.iop.org/1367-2630/11/i=5/a=055060}.

\bibitem[{\citenamefont{Spielman et~al.}(1990)\citenamefont{Spielman, Fesler,
  Eom, Geballe, Fejer, and Kapitulnik}}]{Kerr2}
\bibinfo{author}{\bibfnamefont{S.}~\bibnamefont{Spielman}},
  \bibinfo{author}{\bibfnamefont{K.}~\bibnamefont{Fesler}},
  \bibinfo{author}{\bibfnamefont{C.~B.} \bibnamefont{Eom}},
  \bibinfo{author}{\bibfnamefont{T.~H.} \bibnamefont{Geballe}},
  \bibinfo{author}{\bibfnamefont{M.~M.} \bibnamefont{Fejer}}, \bibnamefont{and}
  \bibinfo{author}{\bibfnamefont{A.}~\bibnamefont{Kapitulnik}},
  \bibinfo{journal}{Phys. Rev. Lett.} \textbf{\bibinfo{volume}{65}},
  \bibinfo{pages}{123} (\bibinfo{year}{1990}),
  \urlprefix\url{https://link.aps.org/doi/10.1103/PhysRevLett.65.123}.

\bibitem[{\citenamefont{Spielman et~al.}(1992)\citenamefont{Spielman, Dodge,
  Lombardo, Eom, Fejer, Geballe, and Kapitulnik}}]{Kerr3}
\bibinfo{author}{\bibfnamefont{S.}~\bibnamefont{Spielman}},
  \bibinfo{author}{\bibfnamefont{J.~S.} \bibnamefont{Dodge}},
  \bibinfo{author}{\bibfnamefont{L.~W.} \bibnamefont{Lombardo}},
  \bibinfo{author}{\bibfnamefont{C.~B.} \bibnamefont{Eom}},
  \bibinfo{author}{\bibfnamefont{M.~M.} \bibnamefont{Fejer}},
  \bibinfo{author}{\bibfnamefont{T.~H.} \bibnamefont{Geballe}},
  \bibnamefont{and}
  \bibinfo{author}{\bibfnamefont{A.}~\bibnamefont{Kapitulnik}},
  \bibinfo{journal}{Phys. Rev. Lett.} \textbf{\bibinfo{volume}{68}},
  \bibinfo{pages}{3472} (\bibinfo{year}{1992}),
  \urlprefix\url{https://link.aps.org/doi/10.1103/PhysRevLett.68.3472}.

\bibitem[{\citenamefont{Xia et~al.}(2006)\citenamefont{Xia, Maeno, Beyersdorf,
  Fejer, and Kapitulnik}}]{Kerr4}
\bibinfo{author}{\bibfnamefont{J.}~\bibnamefont{Xia}},
  \bibinfo{author}{\bibfnamefont{Y.}~\bibnamefont{Maeno}},
  \bibinfo{author}{\bibfnamefont{P.~T.} \bibnamefont{Beyersdorf}},
  \bibinfo{author}{\bibfnamefont{M.~M.} \bibnamefont{Fejer}}, \bibnamefont{and}
  \bibinfo{author}{\bibfnamefont{A.}~\bibnamefont{Kapitulnik}},
  \bibinfo{journal}{Phys. Rev. Lett.} \textbf{\bibinfo{volume}{97}},
  \bibinfo{pages}{167002} (\bibinfo{year}{2006}),
  \urlprefix\url{https://link.aps.org/doi/10.1103/PhysRevLett.97.167002}.

\end{thebibliography}

\end{document}